\documentclass[aps,prl,twocolumn,superscriptaddress,longbibliography]{revtex4-2}

\usepackage{graphicx}
\usepackage{bm}
\usepackage{hyperref}
\usepackage{amsmath}
\usepackage{amssymb}
\usepackage[table]{xcolor}
\usepackage{array}
\usepackage{multirow}
\usepackage[caption=false]{subfig}
\usepackage{physics}

\begin{document}
\title{Band mixing in the quantum  anomalous  Hall regime of twisted semiconductor bilayers}
	\author{Ahmed Abouelkomsan}
  \thanks{ahmed.abouelkomsan@fysik.su.se}
	\affiliation{Department of Physics, Stockholm University, AlbaNova University Center, 106 91 Stockholm, Sweden}
	\affiliation{Department of Physics, Massachusetts Institute of Technology, Cambridge, Massachusetts 02139, USA}

 	\author{Aidan P. Reddy}  \thanks{areddy@mit.edu}
	\affiliation{Department of Physics, Massachusetts Institute of Technology, Cambridge, Massachusetts 02139, USA}

    \author{Liang Fu}\thanks{liangfu@mit.edu}
	\affiliation{Department of Physics, Massachusetts Institute of Technology, Cambridge, Massachusetts 02139, USA}

	\author{Emil J. Bergholtz}\thanks{emil.bergholtz@fysik.su.se}
	\affiliation{Department of Physics, Stockholm University, AlbaNova University Center, 106 91 Stockholm, Sweden}

\date{\today}
\begin{abstract} Remarkable recent experiments have observed fractional quantum anomalous Hall (FQAH) effects at zero field and unusually high temperatures in twisted semiconductor bilayer $t$MoTe$_2$, hence realizing the first genuine fractional Chern insulators. Intriguing observations in these experiments such as the absence of integer Hall effects at twist angles where a fractional Hall effect is observed, do however remain unexplained. The experimental phase diagram as a function of twist angle remains to be established. By comprehensive numerical study, including entanglement spectroscopy, we show that band mixing has large qualitative and quantitative effects on the energetics of competing states and their energy gaps throughout the twist angle range $\theta\leq 4^\circ$. This lays
the foundation for a detailed realistic study of a rich variety of strongly correlated moiré superlattices and an understanding of the phase diagram of these fascinating systems.
\end{abstract}
\maketitle







\textit{Introduction}.
Recent years have witnessed the advent of moir\'e superlattices, made of atomically-thin transition metal dichalcogenides (TMDs), as highly tunable platforms  where experiments have established the existence of a plethora of strongly correlated phases phenomena such as correlated insulators, ferromagnetism and quantum criticality \cite{xu2020correlated,regan2020mott,li2021imaging,huang2021correlated,wangLightinducedFerromagnetismMoire2022,liContinuousMottTransition2021,ghiottoQuantumCriticalityTwisted2021}.

Due to the emergence of narrow flat topological bands, moir\'e superlattices have become an arena for an intricate interplay between topology and correlations. The hallmark example of topological phases is the quantum Hall effect \cite{klitzing1980new,tsui1982two}, both integer and fractional, occurring in Landau levels which arise when two dimensional electrons are subject to strong magnetic fields \cite{RevModPhys.89.025005}. There has been numerous theoretical and experimental research into lattice analogues of the quantum Hall effect in the absence of external magnetic fields, known as the quantum anomalous Hall effect, as these might provide an experimentally feasible way to realize high temperature topological phases \cite{parameswaran2013fractional,bergholtz2013topological,PhysRevLett.106.236802,neupert2011fractional,regnault2011fractional,sheng2011fractional,abouelkomsan2020particle,repellin2020chern,ledwith2020fractional}. Materials exhibiting fractional quantum anomalous Hall effect (FQAH) effect have often been called fractional Chern insulators (FCIs) in the literature \cite{regnault2011fractional,LIU2023}.

The integer quantum anomalous Hall effect (QAH) has been already experimentally observed in different moir\'e systems \cite{li2021quantum,serlin2020intrinsic,taoValleycoherentQuantumAnomalous2022}. In addition, local incompressibility measurements \cite{xie2021fractional} on twisted bilayer graphene point towards the existence of fractional quantum Hall states that survive down to small magnetic fields ($\sim 5 \text{T}$), below which, topologically trivial charge density wave (CDW) states are found instead.

In a very recent exciting development, the first evidence of FQAH states has been observed in twisted $\text{MoTe}_2$ based on both 
thermodynamic and transport measurements \cite{cai2023signatures,zeng2023thermodynamic,park2023observation,PhysRevX.13.031037}. Twisted transition metal dichalcogenide bilayers host topological moir\'e bands with spin/valley contrasting Chern numbers \cite{wu2019topological}. At small twist angles, strong exchange and correlation effects in topological narrow bands \cite{devakul2021magic} drive spontaneous spin/valley ferromagnetism and FQAH states \cite{crepel2023anomalous,li2021spontaneous}.   

Motivated by the current state of art, we provide a detailed study of the many-body interacting problem of AA stacked twisted TMD homobilayers at both integer ($ n = 1$) and fractional ($ n = 1/3$ and $n = 2/3$) hole doping of the underlying moir\'e valence bands. Using unbiased exact diagonalization techniques (ED) that includes multiple bands, we uncover novel significant effects of band mixing on various topological and correlated phases, which also enables a new electronic phase.  

For integer filling $n = 1$, we demonstrate robust QAH states  over a wide range of twist angles. However, for realistic interaction strength at dielectric constant $\epsilon=5$, we find the interacting QAH energy gap as a function of twist angle to be strongly renormalized compared to the non-interacting case, highlighting ubiquitous correlation effects due to band mixing. Remarkably, we also find that strong interactions could drive spontaneous layer polarization (i.e. ferroelectricity) that is only enabled by multiband effects.

Moving to fractional fillings $n = 1/3$ and $n = 2/3$ where previous single-band projected ED studies \cite{li2021spontaneous,reddy2023fractional,wang2023fractional,morales2023pressure,reddy2023toward} have predicted various correlated phases such as FQAH and CDW states, we further show that band mixing plays a crucial role in the stability of the aforementioned phases. In particular, band mixing weakens the FQAH state at $n=1/3$ in favor of the competing CDW state, while at $n=2/3$, FQAH states remain robustly present over a wide range of twist angles. We perform the entanglement spectrum study to characterize and distinguish FQAH and CDW states in finite-size systems.        



\textit{Model}. To model the $\mathbf{K}$ valence bands of twisted bilayer TMDs, we use a continuum description \cite{wu2019topological}. The non-interacting Hamiltonian around a single valley (or spin due to spin-valley locking \cite{zhuGiantSpinorbitinducedSpin2011}) written in the basis of the two layers is given by \begin{equation}
    H_{\uparrow} = \begin{pmatrix}
    \frac{\hbar^2(-i \nabla - \kappa_{+})^2}{2m^*} + V_{+}(\mathbf{r})&& T(\mathbf{r}) \\ T^\dagger(\mathbf{r}) && \frac{\hbar^2(-i \nabla - \kappa_{-})^2}{2m^*} + V_{-}(\mathbf{r})
     \end{pmatrix}
\end{equation}
Where $V_{\pm}(\mathbf{r})$ captures the intralayer moir\'e potential and $T(\mathbf{r})$ represents the interlayer tunneling. Fourier expanding both functions to the lowest harmonics and taking into account symmetry constraints restrict their form to take \begin{equation}
\begin{aligned}
V_{\pm} = 2V \sum_{i = 1,3,5} \cos(\mathbf{G}_i \cdot \mathbf{r})\pm \phi)\\
T(\mathbf{r}) = w ( 1 + e^{-i \mathbf{G}_2 \cdot \mathbf{r}} +  e^{-i \mathbf{G}_3 \cdot \mathbf{r}}).
\end{aligned}
\end{equation}
In the Hamiltonian $H_{\uparrow}$, the corners of the moir\'e Brillouin zone are chosen to be $\boldsymbol{\kappa}_{\pm} = \frac{4 \pi}{3 a_M}(-\sqrt{3}/2,\mp 1/2)$ and the moir\'e recirpocal lattice vectors are given by $\mathbf{G}_i = \frac{4 \pi}{\sqrt{3}a_M}(\cos[(i-1)\pi/3],\sin[(i-1)\pi/3])$ for $ i = 1, \cdots,6 $ where $a_M = a_0 / (2\sin(\theta/2))$ is the moir\'e lattice constant for twist angle $\theta$ and $a_0$ is the lattice constant of the monolayer. The Hamiltonian around the opposite valley $H_{\downarrow}$ is related to $H_{\uparrow}$ by a time-reversal transformation. 

Focusing on twisted $\text{MoTe}_2$  (t$\text{MoTe}_2$), we take $a_0 = 3.52 \> \text{\AA}$ and use the following parameters obtained from fitting the continuum model to DFT calculations \cite{reddy2023fractional}, 
\begin{equation}
    (V,w,\phi, m^*) =(11.2 \> \text{meV}, -13.3 \> \text{meV},-91^\circ, 0.62m_e).
\end{equation}
 We are concerned with the problem of interacting holes in the moir\'e valence bands of  $H_{\uparrow(\downarrow)}$. The full momentum space many-body Hamiltonian reads 
\begin{equation*}
        H = H_0 + H_{\text{int}} 
\end{equation*} with
\begin{multline}
\label{eq:fullHam}
    H_0 = \sum_{\mathbf{\mathbf{k}\alpha \sigma}} \epsilon_{\alpha \sigma}(\mathbf{k}) c^\dagger_{\mathbf{k}\alpha \sigma} c_{\mathbf{k}\alpha \sigma} \\
    H_{\text{int}} = \sum_{\substack{\mathbf{k}_1 \mathbf{k}_2 \mathbf{k}_3 \mathbf{k}_4 \\
    \alpha_1 \alpha_2 \alpha_3 \alpha_4 \\
    \sigma_1 \sigma_2}} V^{\alpha_1 \alpha_2 \alpha_3 \alpha_4}_{\mathbf{k}_1 \mathbf{k}_2 \mathbf{k}_3 \mathbf{k_4}; \sigma_1 \sigma_2}  c^\dagger_{\mathbf{k}\alpha_1 \sigma_1} 
    c^\dagger_{\mathbf{k}\alpha_2 \sigma_2} c_{\mathbf{k}\alpha_3 \sigma_2} 
    c_{\mathbf{k}\alpha_4 \sigma_1} 
\end{multline}
where $c^\dagger_{\mathbf{k}\alpha\sigma}(c_{\mathbf{k}\alpha\sigma})$ are creation (annihilation) operators of holes in a Bloch state $\ket{\mathbf{k}\alpha \sigma}$ where $\alpha$ and $\sigma = \uparrow, \downarrow$ are band and spin indices respectively. $\epsilon_{\alpha\sigma}(\mathbf{k})$ is the single-particle energies and $ V^{\alpha_1 \alpha_2 \alpha_3 \alpha_4}_{\mathbf{k}_1 \mathbf{k}_2 \mathbf{k}_3 \mathbf{k_4}; \sigma_1 \sigma_2}  = \bra{\mathbf{k}_1\alpha_1 \sigma_1; \mathbf{k}_2 \alpha_2 \sigma_2} V\ket{\mathbf{k}_4\alpha_4 \sigma_1; \mathbf{k}_3 \alpha_3 \sigma_2}$ are the two-body interaction matrix elements between the different Bloch states. 
Our two-body interaction is the dual-gated screened Coulomb interaction $V(\mathbf{q}) = 2 \pi e^2 \tanh{d_g |\mathbf{q}|}/(\epsilon|\mathbf{q}|)$ for dielectric constant $\epsilon$ which controls the interaction strength. Unless stated otherwise, we choose the distance $d_g$ from the sample to the gates to be $d_g = 5 \> \text{nm}$.

\begin{figure}[t!]
    \centering
    \includegraphics[width = \linewidth]{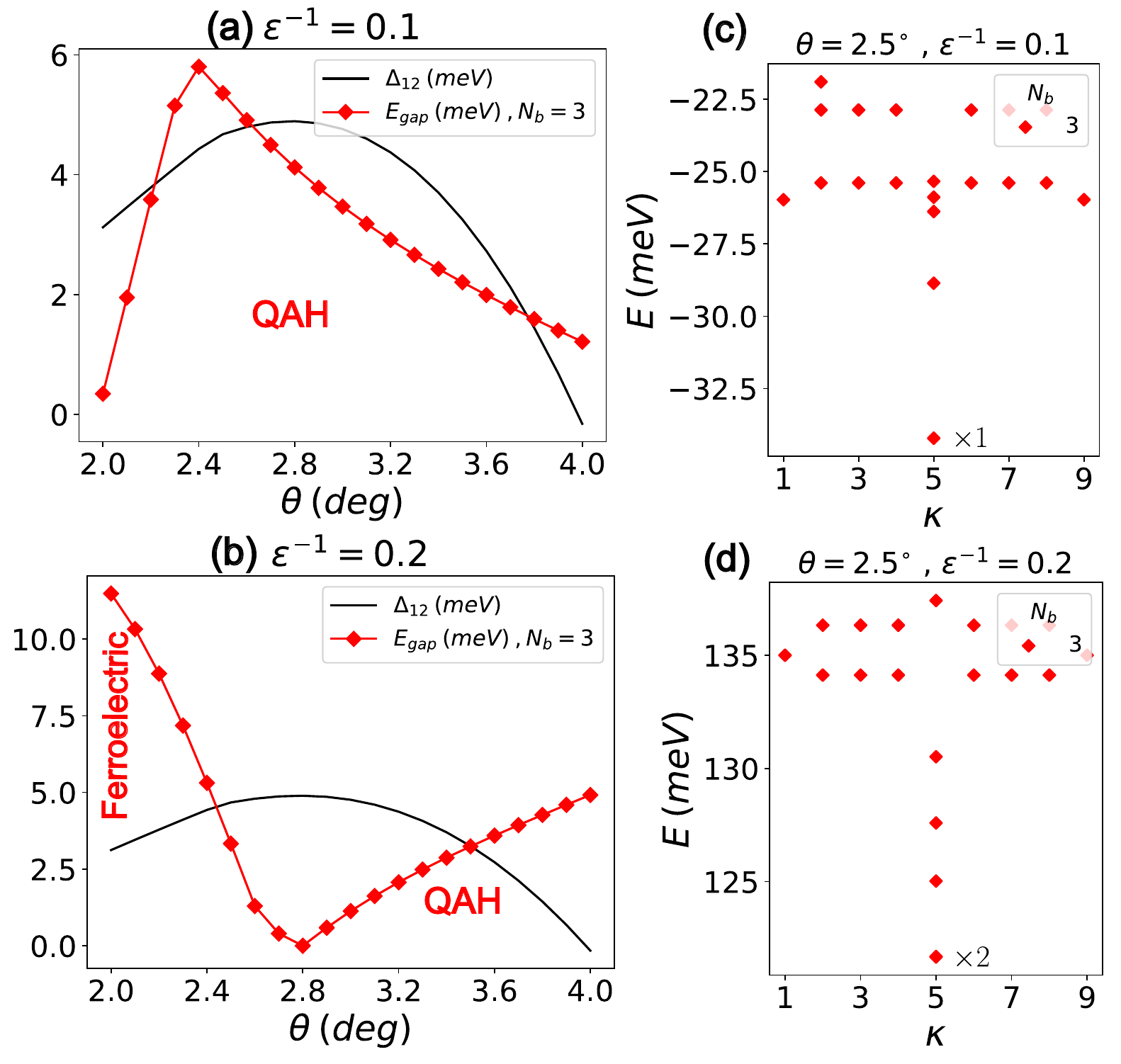}
    \caption{(a)-(b) $E_{\text{gap}}$ at $n = 1$ for (a) $\epsilon^{-1} = 0.1$ where $E_{\text{gap}} = E_2 - E_1$ and (b) $\epsilon^{-1} = 0.2$ where $E_{\text{gap}} = E_3 - E_2$ for $\theta \leq 2.8^\circ$ and $E_{\text{gap}} = E_2 - E_1 $ for $\theta > 2.8^\circ$. $\Delta_{12}$ is the non-interacting band gap between the first two bands. (c)-(d) ED spectrum at $n = 1$ in (c) the quantum anomalous Hall phase  and (d) the ferroelectric phase. Calculations were done on the 9 site cluster \cite{supp} with periodic boundary conditions, $(\theta_1,\theta_2) = (0,0)$.  The index $\kappa$ labels the different momentum points.}
    \label{fig:n_one1}
\end{figure}

We diagonalize the Hamiltonian \eqref{eq:fullHam} on different equal-aspect-ratio clusters \cite{supp} keeping a finite number $N_b$ of valence bands. By the variational principle, our calculations provide an upper bound on the exact ground state energy of Eq. \ref{eq:fullHam} that becomes tighter upon increasing $N_b$. We utilize generic twisted boundary conditions, parameterized by $(\theta_1 \in [0,2 \pi),\theta_2 \in [0,2 \pi))$  along the two axes 
of the cluster. Robust ferromagnetism across a broad range of filling factors $n \leq 1$ has been observed experimentally \cite{cai2023signatures,zeng2023thermodynamic} and in previous numerical studies \cite{reddy2023fractional,crepel2023anomalous}. With this motivation, we perform all calculations in the full spin-valley sector. However, the presence of full spin/valley polarization in ground and low-lying excited states throughout the entire twist angle range studied in this work should not be taken for granted and requires further study.


\begin{figure}[t!]
    \centering
    \includegraphics[width = \linewidth]{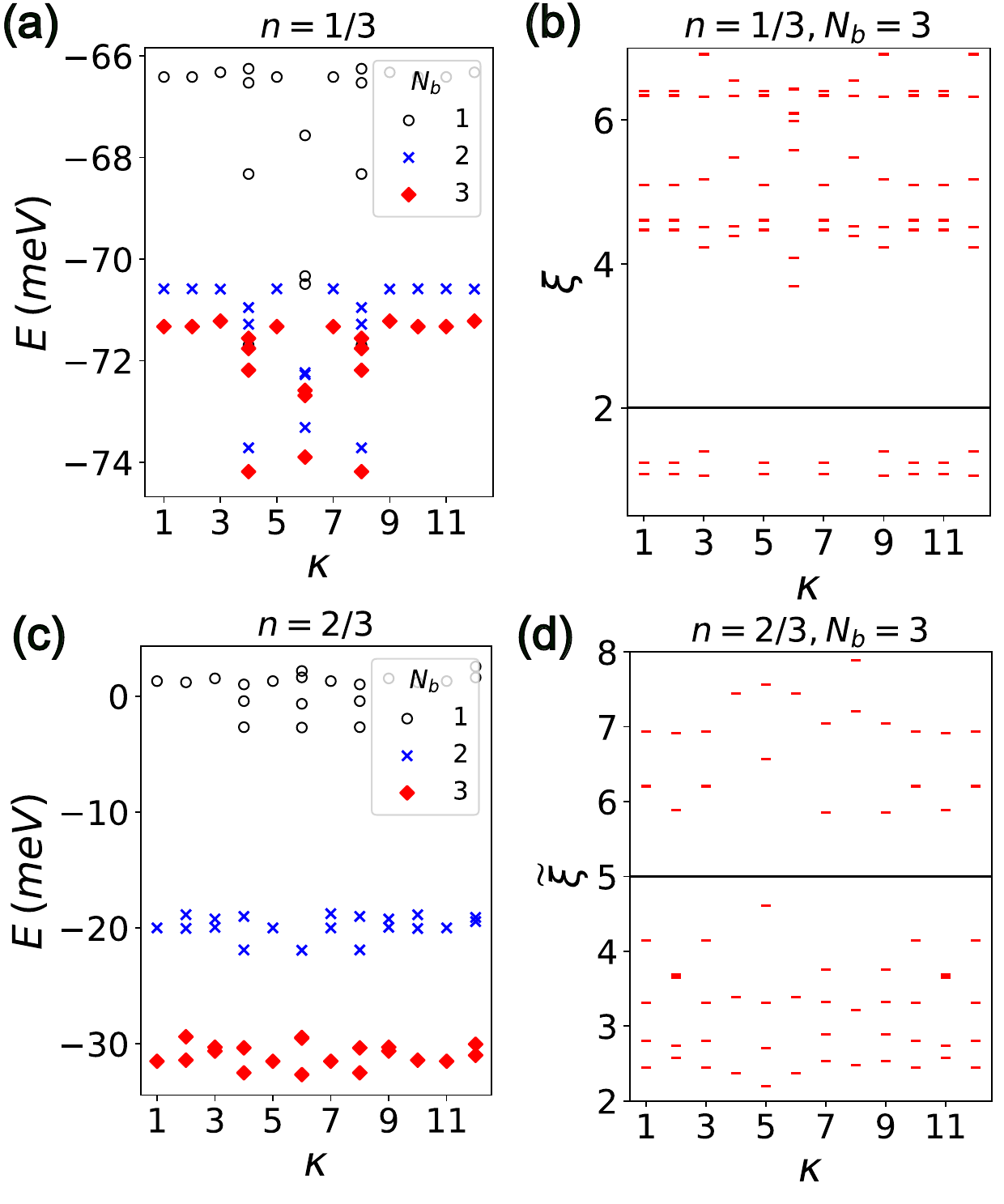}
    \caption{ED results at twist angle $\theta = 2.5^\circ$ and $\epsilon^{-1} = 0.2$. (a) and (c) show the many-body spectrum in the CDW and FQAH phases respectively while (b) and (d) show the PES and the projected (onto the first valence band) PES respectively obtained by keeping $N_A = 2$ particles. There are 18 states and 42 states below the line in (b) and (d), consistent with a CDW and a FQAH state respectively. Calculations were done on the 12 site cluster \cite{supp} using periodic boundary conditions $(\theta_1,\theta_2) = (0,0)$ in (a)-(b) and anti-periodic boundary conditions, $(\theta_1,\theta_2) = (\pi,\pi)$ in (c)-(d). The index $\kappa$ labels the different momentum points.}
    \label{fig:theta2.5}
\end{figure}

\textit{Results}. We begin our analysis by investigating integer hole filling $n = 1$ where previous studies have mainly used self-consistent Hartree-Fock methods \cite{wang2023topological,qiu2023interaction}. For weak interaction strength $\epsilon^{-1} = 0.1$, we find the many-body ground state to be a QAH state for twist angles spanning from $\theta = 2.0^\circ$ to $\theta = 4.0^\circ$. The interacting QAH state is smoothly connected to the non-interacting limit with 
holes completely filling the first valence band which has a Chern number $|C| = 1$. The many-body spectrum shown in Fig \ref{fig:n_one1}(c) at $\theta = 2.5^\circ$ features a single many-body ground state at total momentum $K =  [0,0] $ as expected from a Slater determinant state $\ket{\psi}_{\text{GS}} = \Pi_{\mathbf{k}_i} c^\dagger_{\mathbf{k}\alpha \sigma}  \ket{0}$ for a certain band $\alpha$ and spin $\sigma$. 
However, as shown in Fig \ref{fig:n_one1}(a), the energy gap of the interacting QAH state $E_\text{gap} = E_2 - E_1$ generically differs from the non-interacting energy gap and can be either smaller or larger depending on the twist angle.


As the interaction strength is increased to the realistic value $\epsilon^{-1}=0.2$, we observe the appearance of a ferroelectric phase which exhibits strong layer polarization. The signs of ferroelectricity stem from the existence of two-degenerate many-body ground states (Fig. \ref{fig:n_one1}(d)) which are manifestation of spontaneously breaking $C_{2y}$ symmetry and realize two layer-polarized states on either the top or the bottom layer. 

The ferroelectric phase dominates over the QAH for smaller twist angles. In Fig. \ref{fig:n_one1}(b), we plot $E_{\text{gap}} = E_3-E_2$ as a function of twist angle for realistic interaction strength $\epsilon^{-1} = 0.2$ and find that it is non-vanishing up until $\theta \approx 2.8^\circ$, after which, the system undergoes a transition to a QAH state.

Similar to \cite{abouelkomsan2022multiferroicity}. which focused on smaller twist angles, the emergence of layer polarization can be intuitvely understood from a real-space picture also in the regime that we consider. Here, the first three valence bands have a total Chern number $C = 0$ and admit a real-space description in terms of three Wannier orbitals which are maximally localized on the high symmetry positions, $R^M_X$, $R^X_M$ and $R^M_M$ forming a triangular lattice of three sublattices \cite{qiu2023interaction}. $R^\alpha_\beta$ denotes atomic positions in the moir\'e unit cell where the $\alpha$ atom (metal $M$ or chalcogen $X$) of the top layer is aligned with the $\beta$ atom. The two Wannier orbitals centered on the $R^M_X$ and $R_X^M$ sites are mainly localized in the top and bottom layers respectively while the orbital at the $R^M_M$ site carries equal 
weight in both layers. When the interaction strength is strong, it becomes energetically favorable for the holes to minimize repulsion by localizing on one of the two layer-polarized sublattices. 

The emergence of ferroelectric phase strongly affects the phase space of QAH state. For realistic interaction strength $\epsilon^{-1}=0.2$, the QAH state now appears at $\theta>2.8^\circ$. Interestingly, its topological gap (within the fully polarized sector) increases monotonously (see Fig. \ref{fig:n_one1}(b)) with the twist angle, at least up to $\theta=4^\circ$. 
This contrasts sharply with the noninteracting case where the band gap $\Delta_{12}$ decreases in this angle range.
For comparison, we also performed ED calculation using a different set of continuum model parameters in the literature \cite{wang2023fractional}, and found that at  the QAH phase $n=1$ only appears in a narrow twist angle range for $\epsilon^{-1}=0.1$, and is entirely absent for $\epsilon^{-1}=0.2$ \cite{supp}.  



\begin{figure}[t!]
    \centering
    \includegraphics[width = \linewidth]{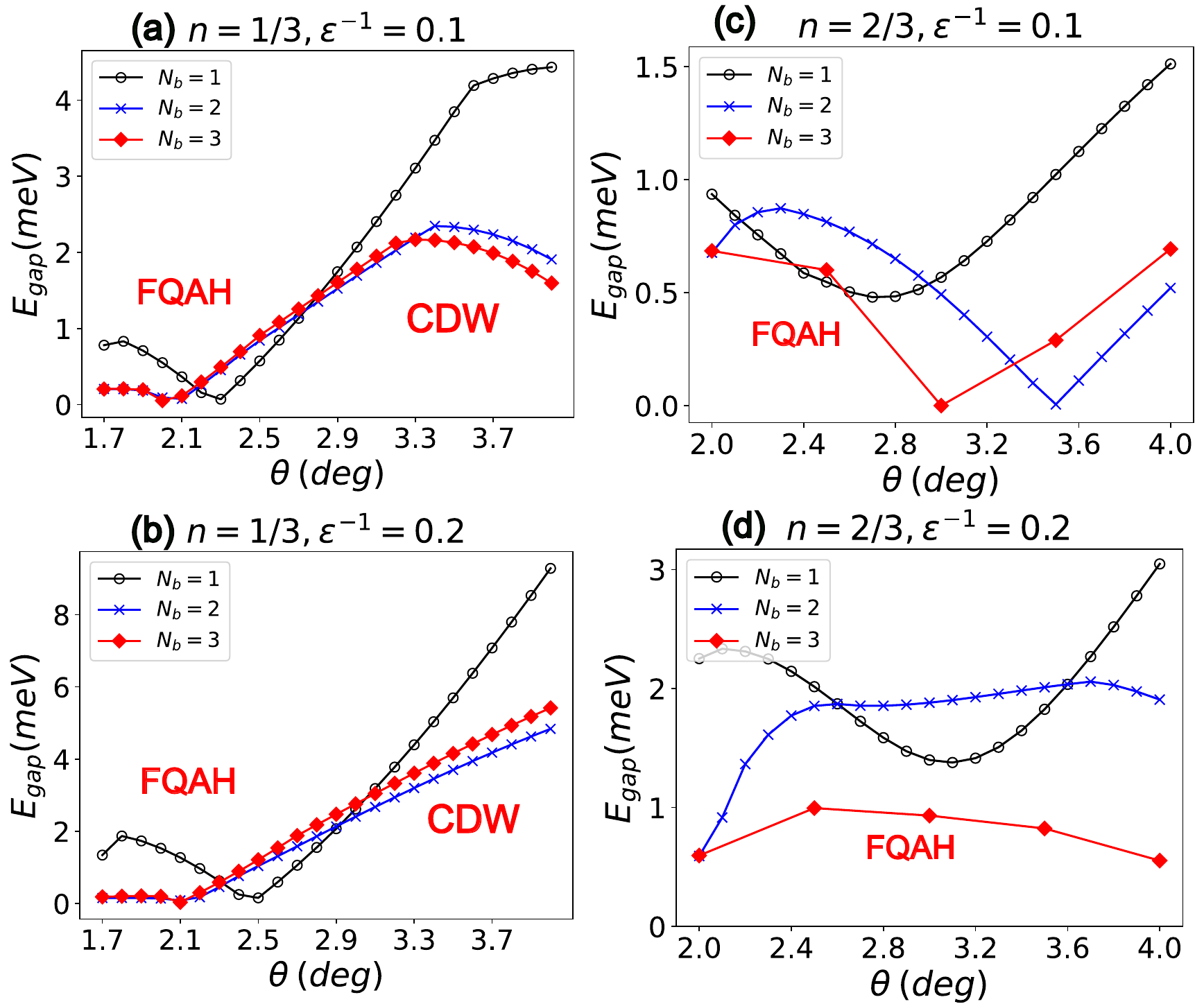}
    \caption{$E_\text{gap} = E_4 - E_3$ as a function of twist angle $\theta$ at $n = 1/3$ and $ n = 2/3$ for $\epsilon^{-1} = 0.1$ and $\epsilon^{-1} = 0.2$. Calculations were done on the 12 site cluster \cite{supp} using periodic boundary conditions $(\theta_1,\theta_2) = (0,0)$ in (a)-(b) and anti-periodic boundary conditions, $(\theta_1,\theta_2) = (\pi,\pi)$ in (c)-(d).}
    \label{fig:gapvstheta}
\end{figure}
Next, we consider fractional hole fillings $n = 1/3$ and $n = 2/3$. For $n = 1/3$, single-band exact diagonalization studies \cite{reddy2023fractional, morales2023pressure, reddy2023toward} have revealed an intriguing interplay between FQAH and CDW states as a function of twist angle while more robust FQAH states for a wider range of twist angles were found at $n = 2/3$. 

To contrast the two fillings, we focus first on twist angle $\theta = 2.5^\circ$ and $\epsilon^{-1} = 0.2$. As shown in Fig. \ref{fig:theta2.5}(a) and  \ref{fig:theta2.5}(c), the many-body spectrum in both cases displays three quasi-degenerate ground states at three distinct total momentum sectors. In this cluster geometry, the ground state momentum sectors expected of FQAH and CDW states happen to be identical and, therefore, distinguishing between the two candidate phases requires further analysis.

In order to pin-point the underlying phase, we calculate the particle entanglement spectrum (PES) \cite{li2008entanglement,sterdyniak2011extracting} of the quasi-degenerate states. The density matrix is defined as $\rho = \frac{1}{3} \sum_{i = 1,2,3} \ket{\Psi_i}\bra{\Psi_i}$ where $\ket{\Psi_i}$ ($i = 1,2,3$) denotes the three quasi-degenerate ground states. We perform a cut in the particle-space corresponding to \textit{minority} particles in a single band, holes for $n = 1/3$ and electrons for $n = 2/3$. By tracing out $N_B$ particles and keeping $N_A$ ones, the entanglement spectrum consists of the eigenvalues $\{\xi_i\}$ of $\text{exp}(-\rho_A)$ where $\rho_A = \text{Tr}_{N_B} \rho$ is the reduced density matrix.  In order to trace out electrons from the first valence band and avoid additional entanglement structure from the filled higher bands, we calculate the entanglement spectrum $\{\widetilde{\xi}_i\}$ obtained from projecting the density matrix $\rho$ onto the first valence band.

As evident from Fig. \ref{fig:theta2.5}(b) and \ref{fig:theta2.5}(d), the PES exhibits a well-separated low-lying spectrum at both fillings. For $n = 2/3$, we find  the number of low-lying states to be consistent with a Laughlin-like FQAH which satisfies a counting rule (1,3) of the number of admissible configurations (at most 1 particle in each 3 consecutive orbitals \cite{haldane1991fractional,bernevig2008model,bergholtz2008quantum}). In contrast, the counting of the fewer low-lying states at $n = 1/3$ is consistent with a CDW state \cite{bernevig2012thin}.

Moreover, we find the CDW state at $n = 1/3$ and the FQAH state at  $n = 2/3$ to exist for a wide range of twist angles. In Fig. \ref{fig:gapvstheta}, we plot the energy gap $E_{\text{gap}} = E_4 - E_3$ as a function of the twist angle $\theta$ for both $\epsilon^{-1} = 0.1$ and $\epsilon^{-1} = 0.2$. In addition to signatures of a weak FQAH state at $n = 1/3$ for $\theta < 2.0^\circ$ \cite{supp}, we observe robust CDW and FQAH states  at fillings $ n = 1/3$ and $ n = 2/3$ respectively for $\theta \geq 2.0^\circ$, using the realistic interaction strength $\epsilon^{-1} = 0.2$.  
For weak interaction $\epsilon^{-1}=0.1$, we find the system at $n = 2/3$ to become metallic for  $\theta \geq 3.5^\circ$.

\begin{figure}[t!]
    \centering
    \includegraphics[width = \linewidth]{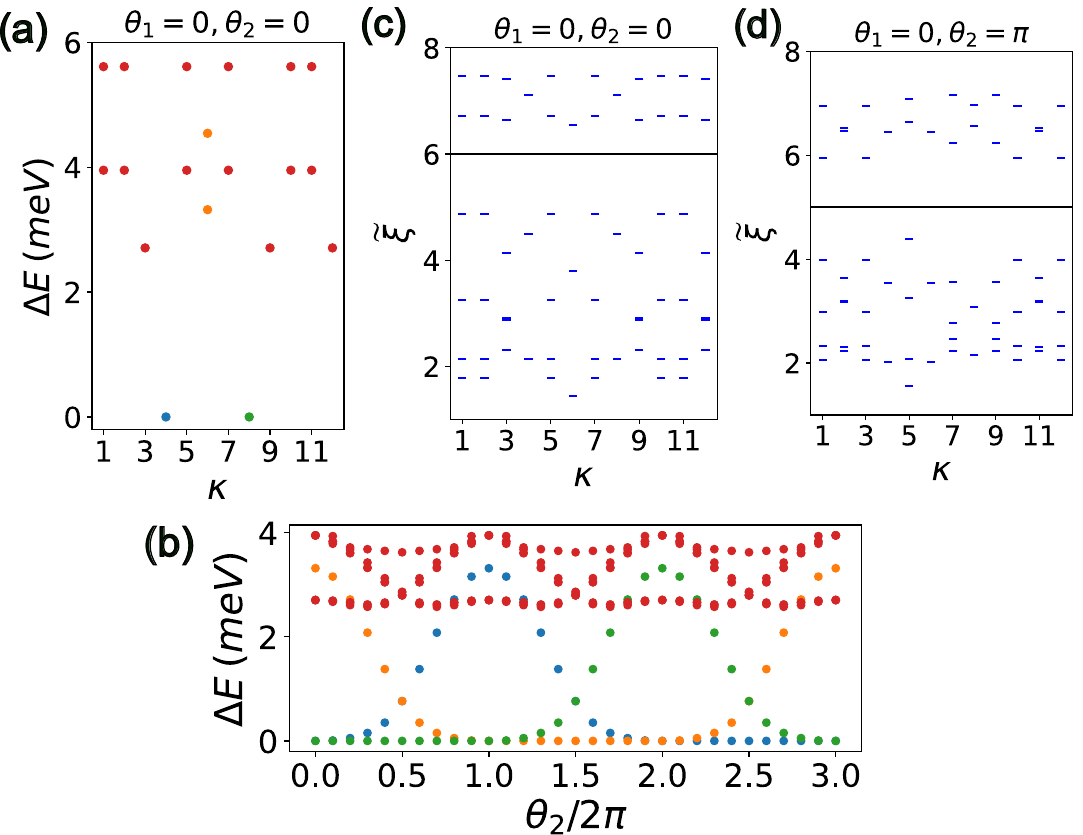}
   \caption{Effect of twisting the boundary conditions for $\theta = 3.6^\circ$, $n = 2/3$ , $N_b = 2$ and $\epsilon^{-1} = 0.2$. (a) The many-body spectrum with periodic boundary conditions, $(\theta_1,\theta_2) = (0,0)$. (b) The evolution of the many-body spectrum upon twisting the boundary condition in one direction ($\theta_2$) and keeping the other direction unchanged ($\theta_1 = 0$). The colors blue, green and orange label the three different momentum sectors where FQAH are expected while the color red labels the rest of the sectors. (c) The projected particle entanglement spectrum at $(\theta_1,\theta_2) = (0,0)$ calculated from the two quasi-degenerate ground states (see (a)). (d)  The projected particle entanglement spectrum at $(\theta_1,\theta_2) = (0,\pi)$ calculated from the three quasi-degenerate ground states.  $N_A = 2$ particles are kept in (c) and (d). There are 42 states below the line in (c) and (d), consistent with a FQAH state.}
    \label{fig:fluxplot}
\end{figure}

We also study the stability of the discussed phases for various twisted boundary conditions. While, we find the phases at $n = 1$ and $ n = 1/3$ to be insensitive to the choice of boundary conditions, we observe significant effects of twisting the boundary conditions at $n=2/3$, see Fig. \ref{fig:fluxplot} where we fix $\theta=3.6^o$. Although FQAH states should be insensitive to changes in the boundary conditions in the thermodynamic limit, these states may exhibit a strong dependence in the small systems available to exact diagonalization. At zero flux (standard periodic boundary conditions) we find an apparent two-fold ground state degeneracy, seemingly at odds with FQAH expectations, see Fig. \ref{fig:fluxplot}(a). 
Crucially, however, we notice that a three-fold set of states originating from the predicted FQAH momenta transform into each other under twisting boundary conditions while evolving essentially separately from the rest of the spectrum as shown in Fig. \ref{fig:fluxplot}(b). Moreover, with anti-periodic boundary conditions in one direction ($\theta_2=\pi$), these three states are separated from the rest of the spectrum and their corresponding PES shows a well developed entanglement gap with the predicted FQAH counting below the gap as shown in Fig. \ref{fig:fluxplot}(d). In fact, this non-trivial counting persists if we consider the PES resulting only the two low lying states at $(\theta_1,\theta_2) = (0,0)$ as evident from Fig. \ref{fig:fluxplot}(c). We take this as strong evidence for a $n=2/3$ FQAH state in the large system limit.

\textit{Discussion}.
Inspired by recent experiments we have shown that multiband effects are crucial for understanding integer and fractional QAH states as well as their competitors at $n\leq 1$ in  twisted semiconductor bilayer $t$MoTe$_2$.

Our results have several important implications. First, it follows that the optimum twist angle for QAH states is filling dependent, i.e. there is no unique magic angle for all fillings. For instance, at twist angles $\theta\lesssim 2.8^\circ$, an integer QAH states is missing at $n=1$ (Fig.~\ref{fig:n_one1}(b)) while a fractional FAQH state at $n=2/3$ prevails (Fig.~\ref{fig:gapvstheta}(d)). Second, a new intriguing phase is enabled by band mixing: spontaneous layer polarized state at $n=1$ (Fig.~\ref{fig:n_one1}(b),(d)). Third, in addition to the particle-hole symmetry breaking within a band, multiband effects provide a second key ingredient in understanding why twisted bilayer $t$MoTe$_2$ may exhibit the FAQH effect at $n=2/3$ but not at $n=1/3$ (Figs.~\ref{fig:theta2.5} and \ref{fig:gapvstheta}). Fourth, the theoretical multiband effects uncovered here, and their relation to experimental results provide important means of distinguishing between the greatly varying available sets of model parameters obtained via first principles calculations. 

 We expect that these effects of multiband mixing will carry over {\it mutatis mutandis} to related setups. This thus provides key input for future studies of a rich variety of strongly correlated twisted semiconductor multilayers. 


\textit{Note added}. During the preparation of this work, a related study of twisted bilayer MoTe$_2$ reported two-band ED results at $n=2/3$ \cite{xu2023maximally}. Upon submission a related preprint appeared \cite{Yu2023}.

  \acknowledgements
	\textit{Acknowledgements}.
A.A. and E.J.B. were supported by the Swedish Research Council (VR, grant 2018-00313), the Wallenberg Academy Fellows program (2018.0460) and and the G\"oran Gustafsson Foundation for Research in Natural Sciences and Medicine. A.A. is also supported by the Wallenberg scholarship program (2022.0348). The work at Massachusetts Institute of Technology is supported by the U.S. Army DEVCOM ARL Army Research Office through the MIT Institute for Soldier Nanotechnologies 
under Cooperative Agreement number W911NF-23-2-0121 and the Simons Foundation.

\bibliography{ref}

\newpage
	\begin{widetext}
 		\renewcommand{\theequation}{S\arabic{equation}}
		\setcounter{equation}{0}
		\renewcommand{\thefigure}{S\arabic{figure}}
		\setcounter{figure}{0}
		\renewcommand{\thetable}{S\arabic{table}}
		\setcounter{table}{0}
 		\section{Supplemental Material}
\section{Exact Diagonalization}
In this paper, we employ exact diagonalization techniques on finite clusters. Let $\mathbf{a}_1$ and $\mathbf{a}_2$ be two basis vectors for the moir\'e superlattice. The finite cluster is spanned by two vectors $\mathbf{T}_1$ and $\mathbf{T}_2$ given by 
\begin{equation}
\begin{aligned}
    \mathbf{T}_1 = m_1 \mathbf{a}_1 + n_1 \mathbf{a}_2 \\
\mathbf{T}_2 = m_2 \mathbf{a}_1 + n_2 \mathbf{a}_2
\end{aligned}
\end{equation}
where $m_1$, $n_1$, $m_2$ and $n_2$ are integers. Generic twisted boundary conditions are imposed by requiring $\mathbb{T}_{\mathbf{T}_1} \psi_{\mathbf{k}\alpha\sigma}({\mathbf{r}}) = e^{i \theta_1}\psi_{\mathbf{k}\alpha\sigma}({\mathbf{r}})  $ and $\mathbb{T}_{\mathbf{T}_2} \psi_{\mathbf{k}\alpha\sigma}({\mathbf{r}}) = e^{i \theta_2}\psi_{\mathbf{k}\alpha\sigma}({\mathbf{r}})  $ where $\mathbb{T}$ is the translation operator and $\psi_{\mathbf{k}\alpha\sigma}({\mathbf{r}})$ is a Bloch state of  momentum $\mathbf{k}$, band $\alpha$ and spin $\sigma$.

In momentum space, this amounts to a discrete set of momentum points given by $\mathbf{k} = (\alpha + \theta_1/ 2 \pi) \> \mathbf{g}_1 + (\beta + \theta_2/ 2 \pi) \> \mathbf{g}_2$ for integer $\alpha$ and $\beta$ where $\mathbf{g}_1$ and $\mathbf{g}_2$ are the discretization vectors of the momentum grid, given by $\mathbf{g}_1 = 1/(m_1 n_2 - n_1 m_2)(n_2 \mathbf{G}_1  - m_2 \mathbf{G}_2$) and $\mathbf{g}_2 = 1/(m_1 n_2 - n_1 m_2)(-n_1 \mathbf{G}_1  + m_1 \mathbf{G}_2$) with $\mathbf{G}_1$ and $\mathbf{G}_2$ are the recpriocal lattice basis vectors. In the main text, we use a 9 site cluster and a 12 site cluster shown in Fig \ref{fig:clusterplot} 

\begin{figure}[h!]
    \centering
    \includegraphics[width = 0.5\linewidth]{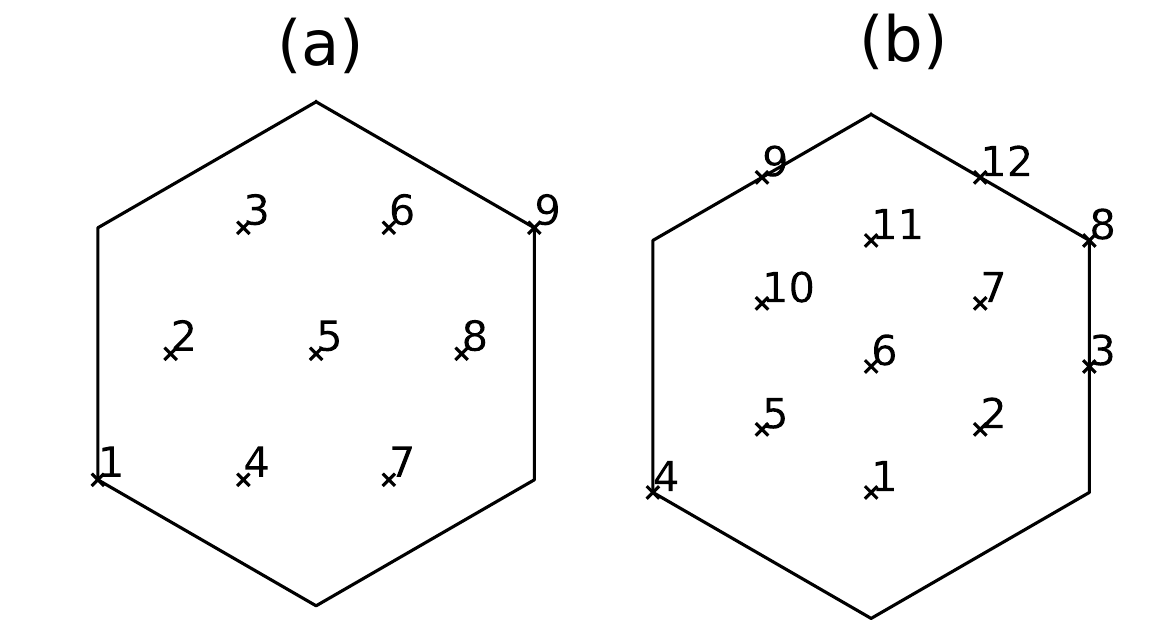}
   \caption{Momentum points of the (a) 9 site cluster and (b) 12 site cluster for $(\theta_1, \theta_2) = (0,0)$}
    \label{fig:clusterplot}
\end{figure}

  \section{Particle Entanglement Spectrum at $n = 1/3$, $\theta < 2.0 ^\circ$}
  Here, we provide evidence for a FQAH state at $n = 1/3$ for twist angles $\theta \leq 2.0^\circ$ (c.f Fig \ref{fig:gapvstheta}(a)-(b)) from particle entanglement spectroscopy. We take $\theta = 1.9 ^\circ$ as an example. As shown in Fig \ref{fig:PESsupp}, the PES has a spectral gap, below which the number of states matches the counting expected from a Laughlin type FQAH. This stands in constrast to the counting in the CDW phase, shown in Fig. \ref{fig:theta2.5}(b) which has fewer states.
  \begin{figure}[h!]
    \centering
    \includegraphics[width = 0.5\linewidth]{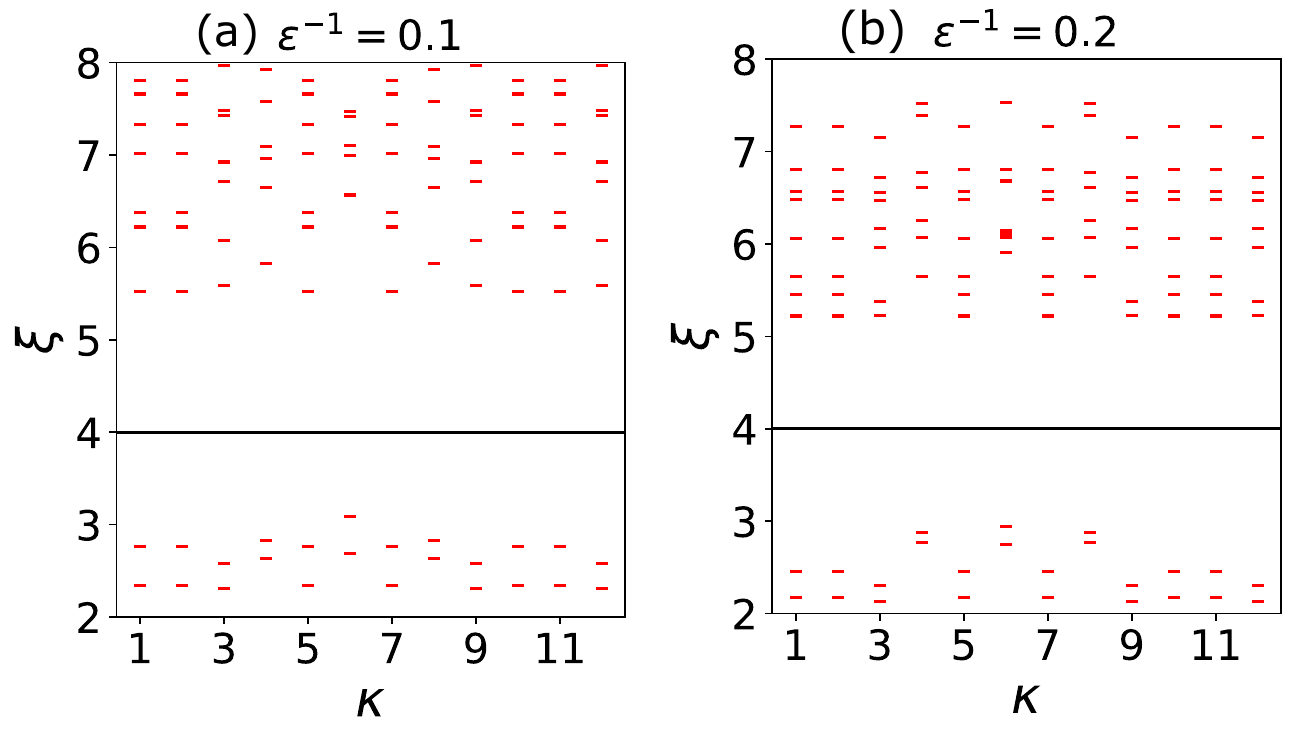}
   \caption{PES calculated at $\theta = 1.9^\circ$, $n = 1/3$ and $N_b = 3$ for (a) $\epsilon^{-1} = 0.1$ and (b) $\epsilon^{-1} = 0.2$. $N_A = 2$ particles are kept. There are 42 states below the line, consistent with a FQAH state. }
    \label{fig:PESsupp}
\end{figure}
  \section{Results using Ref \cite{wang2023fractional} parameters}
  In this sections, we study the continuum model of  $t$MoTe$_2$ using the parameters introduced in Ref \cite{wang2023fractional}. We focus at filling $n = 1$.  As shown in Fig. \ref{fig:UWsupp}, we find signs of QAH states only for weak interaction $\epsilon^{-1} = 0.1$ at twist angles $\theta > 3.5^\circ$ while we find robust ferroelectricity for $\theta \leq 3.5 ^\circ$ and for the realistic interaction $\epsilon^{-1} = 0.2$ at all twist angles between $\theta = 3.0^\circ$ and  $\theta = 4.0^\circ$.
    \begin{figure}[h!]
    \centering
    \includegraphics[width = 0.8\linewidth]{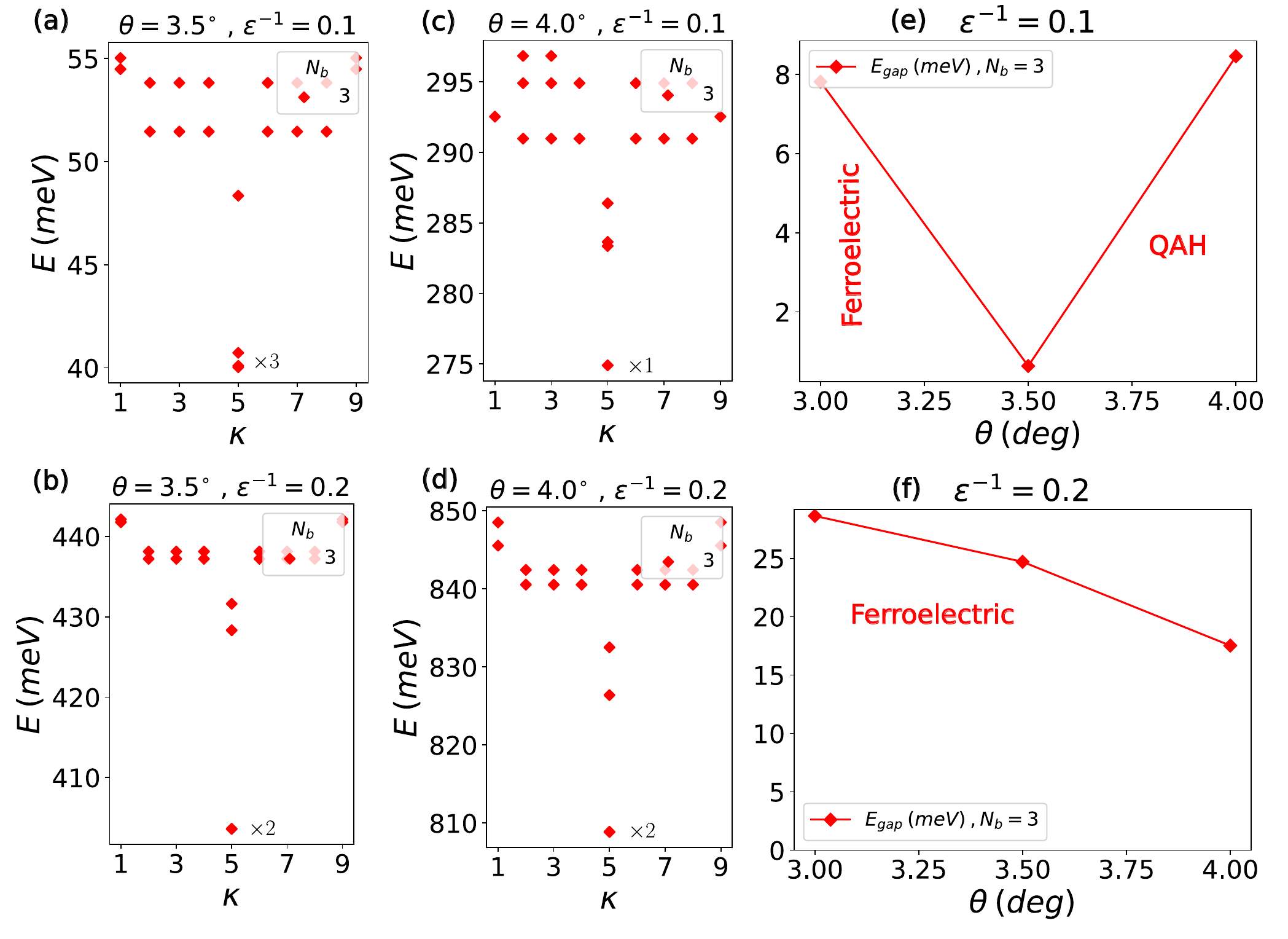}
   \caption{(a)-(d) The many body spectrum on the 9 site cluster \ref{fig:clusterplot}(a) using Ref \cite{wang2023fractional} parameters at $n =1$ and $N_b = 3$. (e)-(f) Energy gap $E_{\text{gap}}$ as a function of twist angle $\theta$ where in (e), $E_{\text{gap}} = E_3-E_2$ for $\theta \leq 3.5 ^\circ$ and  $E_{\text{gap}} = E_2-E_1$ for $\theta > 3.5 ^\circ$ while  $E_{\text{gap}} = E_3-E_2$ in (f).}
    \label{fig:UWsupp}
\end{figure}

\end{widetext}

\end{document}